# Set Shaping Theory and the Foundations of Redundancy-Free Testable Codes

Aida Koch, Alix Petit

Abstract: To render a sequence testable, namely capable of identifying and detecting errors, it is necessary to apply a transformation that increases its length by introducing statistical dependence among symbols, as commonly exemplified by the addition of parity bits. However, since the decoder does not have prior knowledge of the original symbols, it must treat the artificially introduced symbols as if they were independent. Consequently, these additional symbols must be transmitted, even though their conditional probability, under ideal and error-free conditions, would be zero. This sequence extension implies that not all symbol combinations of the new length are practically realizable: if an error modifies a sequence, making it "inadmissible," such an error becomes detectable. Recent developments from Set Shaping Theory have demonstrated a significant result: there always exists a subset of dimension $A^N$ of sequences with an extended length ($N_2 > N$) having an alphabet A, whose average value of zero order empirical $N_2H_0(S2)$ is slightly lower than the average value of zero order empirical $NH_0(S)$ of the sequences S belonging to the set of size $A^N$ composed of all sequences of length N and alphabet A. This finding implies that a sequence S of length N having an alphabet A can be transformed in a biunivocal way into a sequence S2 of length N2 can be encoded as if composed of independent symbols, achieving the desired testability without explicitly introducing redundancy.

## Introduction

The reliability and accuracy of digital communication systems critically depend on the ability to detect and correct errors that may occur during data transmission [1]. Traditional approaches to error detection involve introducing redundancy by increasing the sequence length and embedding statistical dependencies among symbols, such as the widely used parity bit method [2]. However, this redundancy often comes with the drawback of reduced transmission efficiency, as the decoder, lacking prior knowledge of the original symbols, must interpret the additional symbols as independent, thereby requiring their explicit transmission even when their conditional probability ideally would be zero.

Set Shaping Theory (SST), an emerging field in information theory, recently introduced a result of great significance: there always exists a subset of dimension $A^N$ of sequences with an extended length ($N_2 > N$) having an alphabet A, whose average value of zero order empirical $N_2H_0(S2)$ is slightly lower than the average value of zero order empirical $NH_0(S)$ of the sequences S belonging to the set of size $A^N$ composed of all sequences of length N and alphabet A [3]. This means that it's possible to use extended sequences to detect errors while maintaining the appearance of independence without explicitly introducing redundant bits by selecting only valid sequences for transmission.

Contact author:  aida.koch445@outlook.com

## Testable Codes: Motivation and Foundations

In the design of reliable communication systems and storage devices, it is often essential not only to **correct** errors but also to **detect** their presence efficiently. This has led to the development of **testable codes**, which are coding schemes designed to expose deviations from valid codewords through structured statistical or algebraic properties [4]. A code is said to be *testable* if it allows, with high probability, the detection of errors through local or global checks—often without needing full decoding. This capability is crucial in environments where errors are rare but costly, and fast detection is required before initiating more complex correction procedures [5].

Testable codes often rely on **structured redundancy**, where specific dependencies are imposed among symbols in the codeword. A classic example is the parity bit, where the last symbol is a deterministic function of the others. However, more advanced constructions introduce global constraints that define a subset of sequences with particular properties, such as low-density parity-check (LDPC) codes or algebraic codes like Reed–Solomon codes. In such settings, a received sequence not belonging to the valid code set can be flagged as erroneous, often using only partial information.

From an information-theoretic point of view, the introduction of such redundancy appears to reduce the effective information rate. However, recent advances, including the **Set Shaping Theory**, suggest that one can construct codebooks of shaped sequences that maintain error detectability without necessarily increasing entropy, offering a novel balance between redundancy and efficiency [6].

## Set Shaping Theory: An Overview

Set Shaping Theory (SST) represents a modern approach in information theory that goes beyond Shannon's classical framework. Rather than passively accepting the entropy limits of sequences, SST actively reshapes datasets to reduce their effective information content via a structured transformation.

Let S be a string of length N over the alphabet $A = \{a_{i,\ldots\ldots\ldots\ldots}a_h\}$, and let $n_i$ denote the number of occurrences of the symbol $a_i$ inside S. The zeroth order empirical entropy of the string S is defined as:

$$H0(\boldsymbol{S}) = -\sum_{i=1}^{h} n_i/n \log_2 n_i/n$$

The zero-order empirical entropy $H_0(S)$ multiplied by the length of the sequence N is considered to be coding limit of an individual sequence through the fixed code symbols using a uniquely decipherable and instantaneous code.

Contact author: aida.koch445@outlook.com

The Set Shaping Theory has as its objective the study and application in information theory of bijection functions $f$ that transform a set $X^N$ of strings of length $N$ into a set $Y^{N+K}$ of strings of length $N+K$ with $K$ and $N \in \mathbb{N}^+$, $|X^N| = |Y^{N+k}|$ and $Y^{N+K} \subset X^{N+K}$.

The function $f$ defines from the set $X^{N+K}$ a subset of size equal to $|X^N|$. This operation is called "Shaping of the source", because what is done is to make null the probability of generating some sequences belonging to the set $X^{N+K}$.

The parameter $K$ is called the shaping order of the source and represents the difference in length between the sequences belonging to $X^N$ and the transformed sequences belonging to $Y^{N+k}$.

The functions that respect this condition are many but since the goal of this theory is the transmission of data, the function $f_m$ studied is which transforms the set $X^N$ into the set $Y^{N+k}$ composed of the $|X^N|$ strings with less $H_0(x)$ belonging to $X^{N+k}$.

The bijection function $f_m$ is defined as:
$$f_m: X^N \to Y^{N+k}$$
With $K, N \in \mathbb{N}^+$, $|X^N| = |Y^{N+k}|$, $Y^{N+K} \subset X^{N+K}$, $X^{N+K} - Y^{N+K} = C^{N+K}$, $\forall\, y \in Y^{N+K}$ and $\forall\, c \in C^{N+K}$ $H_0(y) < H_0(c)$ and $H_0(y_i) < H_0(y_{i+1})$ $\forall\, y \in Y^{N+K}$.

Given a set $X^N$ which contains all the sequences of length N that can be generated, therefore with dimension $|X^N| = |A|^N$. The Set Shaping Theory tells us that when $|A| > 2$ exists a set $Y^{N+K}$ of dimension $|A|^N$ consisting of sequences having alphabet A and length N+K, in which the average value of $(N+k)H_0(y)$ is less than the average value of $NH_0(x)$ calculated on the sequences belonging to $X^N$.

Since the two sets have the same dimension, it is possible to put the sequences belonging to the two sets into a one-to-one relationship.

## A New Paradigm: Compression Meets Robustness

Error Detection Without Redundancy: In traditional coding, redundancy (parity bits) is necessary for detecting errors. SST creates *forbidden patterns* sequences whose conditional probability is zero. If any such pattern is observed during decoding, an error is signaled without the need for any extra check bits.
This gives rise to a new class of *locally testable codes*, where verification is embedded within the shape of the sequence, not added externally.

By applying SST to message reshaping before encoding, one achieves a twofold gain: lower entropy and automatic error detection. The bijection:
$f: X^N \to Y^{N+K}$,    $|X^N| = |Y^{N+K}| \subset X^{N+K}$
produces output sequences in which certain symbol patterns are structurally inadmissible due to zero conditional probabilities. This means any deviation from the allowed shaped set during

Contact author: aida.koch445@outlook.com

transmission is inherently detectable no parity checks needed.

Thus, the code becomes *testable by design*. The decoder needs only to verify whether a received sequence belongs to the shaped set. If not, an error is flagged.

Importantly, empirical results show that the increase in sequence length is more than compensated for by a decrease in per-symbol information, maintaining (or even improving) overall compression efficiency.

## Implementation Challenges and Research Directions

While SST is promising in theory, its real-world application comes with obstacles:

- Memory and Computation: Mapping functions used in SST require large tables to represent bijections from $X^N$ to $Y^{N+K}$, which become impractical as $N$ grows. Research is ongoing to develop compact and efficient approximations.
- Experimental Results: Empirical studies applying SST to Huffman encoding have validated theoretical predictions, demonstrating entropy reduction and error detection simultaneously [7].
- Beyond Extension: Emerging work is exploring *negative shaping orders* (sequence shortening), as well as broader uses in source modeling, data compression [8].

## Conclusion

Set Shaping Theory introduces a groundbreaking shift in how we think about the development of the testable codes. Instead of adding error-checking bits, SST embeds error detection directly into the structure of the encoded sequence by transforming the sequence before encoding.

To make a sequence *testable* that is, capable of revealing the presence of errors, it is necessary to transform it by **increasing its length** and **introducing dependencies among symbols**, as in the case of parity bits. However, the decoder does not know the original data and must treat the newly introduced (redundant) symbols as if they were independent. This means these symbols must be **explicitly transmitted**, even though their conditional probability given the original data would ideally be zero in the absence of errors.

This transformation increases the sequence length in such a way that **not all combinations of symbols** of the new length are admissible. As a result, if an error alters the sequence, producing a pattern that should not occur under the encoding rule, the system can detect the error.

Recent developments in **Set Shaping Theory** have revealed a surprising result: it is always possible to transform a sequence into a longer version by carefully selecting which longer sequences are allowed, in such a way that the overall set of sequences becomes **more structured** and **less complex** than the original. This means that even though the sequence is extended and dependencies are introduced between symbols, the total amount of information contained in the new set does **not** increase proportionally on the contrary, it can be slightly **reduced**.

In other words, one can construct a new set of longer sequences where each one corresponds uniquely to an original sequence, but the entire set is designed in such a way that it can be treated **as if the symbols were independent**, making encoding simpler. This allows the

Contact author: aida.koch445@outlook.com

sequence to become testable capable of detecting errors **without adding visible redundancy** or increasing the informational burden.

The key implication is that **testability can be achieved without increasing entropy** or introducing explicit redundancy, by selecting and shaping the set of valid output sequences according to the principles of Set Shaping Theory.

This approach unifies two traditionally separate concerns compression and error detection into a single framework. It opens the door to redundancy-free encoding systems that are both efficient and reliable, guided by foundational principles of information theory.

Contact author: aida.koch445@outlook.com